\documentclass{aa}
\usepackage{txfonts}
\usepackage{graphicx}
\usepackage{natbib}
\bibpunct{(}{)}{;}{a}{}{,}
\newcommand{\hei}{He~{\sc i}}
\newcommand{\heii}{He~{\sc ii}}
\newcommand{\Oii}{O {\sc ii}}
\newcommand{\mue}{\mu}
\newcommand{\kms}{$\mathrm{kms^{-1}}$}
\begin{document}


\hyphenation{ana-lysed do-mi-nant}

\title{The MSST Campaign: II.Effective temperature and gravity variations in the multi-periodic 
pulsating subdwarf B star PG1605+072\thanks{Based on observations at the Bok Telescope on Kitt Peak, operated by Steward Observatory (University of Arizona).}\fnmsep\thanks{Based on observations made with the Nordic Optical Telescope, operated
on the island of La Palma jointly by Denmark, Finland, Iceland,
Norway, and Sweden, in the Spanish Observatorio del Roque de los
Muchachos of the Instituto de Astrofisica de Canarias.}\fnmsep\thanks{Based on observations 
collected at the European Southern Observatory, Chile.}}
\author{A. Tillich\inst{1}\and U. Heber\inst{1}\and S. J. O'Toole\inst{2}\and 
R. \O{}stensen\inst{3}\and S. Schuh\inst{4}}
\offprints{A.~Tillich\\
\email{Alfred.Tillich@sternwarte.uni-erlangen.de}}
\institute{Dr.Remeis-Sternwarte Bamberg, Universit\"at Erlangen-N\"urnberg, 
Sternwartstr.7, D-96049 Bamberg, Germany\and Anglo-Australian Observatory, 
P.O. Box 296 Epping, NSW 1710, Australia\and Instituut voor Sterrenkunde, 
Celestijnenlaan 200D, 3001 Leuven, Belgium\and Institut f\"ur Astrophysik, 
Universit\"at G\"ottingen, Friedrich-Hund-Platz 1, 37077 G\"ottingen, Germany }

\date{Received 25 May 2007 / Accepted 28 June 2007}

\abstract{
Stellar oscillations are an important tool to probe the
interior of a star.
Subdwarf B stars are core helium burning objects, but their formation
is poorly understood as neither single star nor binary
evolution can fully explain their observed properties.
Since 1997 an increasing number of sdB stars has been
found to pulsate forming two classes of stars
(the V361 Hya and V1093 Her stars).
}
{
We focus on the bright V 361 Hya star PG1605$+$072 to
characterize its frequency spectrum. While most previous
studies relied on light variations, we have measured
radial velocity variations for as much as 20 modes. In this paper
we aim at characterizing the modes from atmospheric
parameter and radial velocity variations.
}
{
Time resolved spectroscopy
($\approx$9000 spectra) has been carried out to detect
 line profile
variations from which variations of the effective
temperature and gravity are extracted by means of a quantitative
spectral analysis.
}
{
We measured variations of effective temperatures and gravities
for eight modes with semi-amplitudes ranging from
$\Delta T_{\rm{eff}}=880$\,K to as small as 88\,K and $\Delta\log{g}$
of 0.08\,dex to as low as 0.008\,dex.
Gravity and temperature vary almost in phase,
whereas phase lags are found between temperature and
radial velocity.
}
{
This profound analysis of a unique data set serves as sound basis 
for the next step towards an identification of pulsation modes. 
As rotation may play an important role the modelling of pulsation 
modes is challenging but feasible.
}


\titlerunning{MSST spectroscopy of PG1605+072: II.Analysing line profile variations}

\maketitle


\section{Introduction}
The subluminous B stars (commonly referred to as sdBs) are generally
believed to be core helium burning stars with very thin ($<$0.02\,$M_\odot$)
hydrogen envelopes which have masses around
$\sim$0.5\,$M_\odot$. Considerable evidence has accumulated that these stars
are sufficiently common to be the most likely source for the ``UV upturn
phenomenon'' observed in elliptical galaxies and spiral galaxy bulges 
\citep{1997ApJ...486..201Y}. However, important questions still
remain about their formation and the appropriate timescales.  Following ideas 
outlined by \cite{1986A&A...155...33H}, the sdBs
can be identified with models for Extreme Horizontal Branch (EHB) stars. An
EHB star bears great resemblance to a helium main-sequence star of half a
solar mass and it should evolve similarly, i.e. directly to the white dwarf
cooling sequence, bypassing a second giant phase. 

While the next stages of sdB evolution seem to be well
known, the question of the stars' formation is widely unanswered. 
Nowadays evidence is growing that close binary 
evolution plays an important role for the formation of sdB stars 
\citep{2001MNRAS.326.1391M,2003MNRAS.338..752M,2004Ap&SS.291..321N}.

The work of \cite{2003MNRAS.341..669H} brought remarkable progress.
Three possible binary formation channels were studied: common envelope ejection, stable Roche
lobe overflow and the merger of two helium white dwarfs.
One of the key results of the Han et al.\ studies was the suggestion that the
mass range of sdBs may be larger than previously thought. They found that sdB
stars in binary systems may have masses as low as 0.36\,$M_\odot$ and still
pass through a helium core burning phase. 
Until now the mass of any sdB has always been assumed to be
$\sim$0.5\,$M_\odot$ for the determination of luminosity or binary companion
mass, however now it appears this may not be entirely appropriate. What is
needed is a set of precise masses that can be compared with sdB formation
models such as those of Han et al. 


Asteroseismology provides a promising avenue to determine masses of stars.
Two classes of non-radially 
pulsating sdB stars have been discovered recently, 
the V361~Hya and V1093 Her stars \citep{1997MNRAS.285..640K, 2003ApJ...583L..31G}. 
The V~361 Hya ones are of short period (2--8\,min) and
have photometric amplitudes of typically less than 10\,mmag, while
the V1093 Her stars have longer periods (45 to 120\,min) and even smaller 
amplitudes of less than 2\,mmag. Several of the pulsating 
sdB stars reside in close binary systems. 

The V361~Hya stars have already been proven to be extremely useful objects for
asteroseismology, 
because many of them show a sufficient number of modes with only small amplitude variations 
rendering them ideal targets for long-term photometric campaigns to resolve the 
frequencies of the modes in their light curves. Single site monitoring was
carried out for several V~361 Hya stars. However, the complexity of the 
oscillations requires high frequency resolution which can only be achieved in 
extensive multi-site campaigns. Therefore observers have teamed up 
to continously collect photometry over weeks for a few V~361 Hya stars 
\citep[e.g.][]{1999MNRAS.303..525K, 2002A&A...389..180S, 2006A&A...459..557S,
2006ApJ...643.1198R}. 

Theoretical studies by \citet{1997ApJ...483L.123C} showed that the 
oscillations of 
V361 Hya stars can be explained as acoustic modes $(p-modes)$ of low degree $l$ and
low radial order $n$ excited by an opacity bump due to a local enhancement of
iron-group elements. 
Theoretical modelling has now matured to derive
stellar parameters (stellar mass and hydrogen envelope mass) by matching model 
predictions to the observed frequency spectra 
\citep[see ][ and reference therein]{2006A&A...459..565C}.

Light variations provide important information on the oscillation properties. 
In order to identify the pulsation modes reliably, however, 
it is desirable to supplement them with measurements of stellar surface motions 
leading to radial velocity and line profile variations. This has been achieved
for the first time for PG~1605$+$072 
\citep[][ see below]{2000ApJ...537L..53O} and attempted
with diverse success for half a dozen V~361 Hya stars 
\citep{2000MNRAS.318..974S,2002MNRAS.332...34W,2003whdw.conf...95W,2004A&A...419..685T,2006A&A...450.1149T}.

PG1605$+$072 (also known as V338 Ser) was discovered by \citet{1998MNRAS.296..317K} to be a V361\,Hya
and was found to have the longest periods (up to 9\,min) of this class of stars and a very large 
photometric amplitude (up to $\sim$60\,mmag).
They detected a very strong main mode together with about 20 other modes. 
In the following years a first multi-site campaign confirmed these results and 
increased the number of observed frequencies up to 50 \citep{1999MNRAS.303..525K}.
 Using high resolution Keck spectra, the stellar parameters and metal 
 abundances were derived by \citet{1999A&A...348L..25H} . 
 They found $T_{\rm{eff}}=32,300$\,K, 
 $\log{g}=5.25$\,dex and $\log{(He/H)}=-2.53$, whereas the small 
 $\log{g}$ implies that the star is quite evolved and has already moved away 
 from the EHB. 


Radial velocity variations due to pulsation were detected for the first time by
\citet{2000ApJ...537L..53O} and confirmed by
\citet{2002MNRAS.334..471O,2002MNRAS.329..497W,2003A&A...401..289F}. 
Variations of equivalent widths (line indices) were detected by \citet{2003MNRAS.340..856O}
and analysed for $T_{\mathrm{\rm{eff}}}$ ($\Delta T_{\mathrm{eff}}\approx560$\,K) and $\log{g}$ ($\Delta\log{g}\approx0.062$\,dex) variations. 
Nevertheless these studies suffered from poor frequency resolution or from the lack of
simultaneous photometry. Therefore we organised a multi-site coordinated
spectroscopic campaign to observe PG1605$+$072 with medium resolution
spectrographs on 2\,m and 4\,m telescopes \citep[MSST, ][]{2003whdw.conf..105H}. 
Surface motions for more than 20 pulsation modes were detected in this data set
from radial velocity variations \citep[][ henceforth Paper~I]{2005A&A...440..667O}. 
In this paper we proceed to 
analyse the same data set for line profile variation in order to 
search for, analyse, and interpret variations in surface gravity 
and effective temperature.

The paper is organised as follows. In Sect.~\ref{sec:data} we outline the 
available data and in Sect.~\ref{sec:spec_analysis} the 
derivation of atmospheric parameter variations by fitting model atmospheres to 
observed spectral lines. The phase lags between radial velocity, temperature and
gravity variations are measured in Sect.~\ref{sec:phase_lags}, 
leading to a discussion in the last section including an 
outlook to the theoretical modelling of line profile variations.


\section{The MSST 2m spectroscopy}\label{sec:data}

During the Multi-Site Spectroscopic Telescope (MSST) campaign  
we obtained 151 
hours of time-resolved spectroscopy on PG\,1605+072 at four observatories
in May/June 2002 along with extensive time-resolved photometry on 2\,m class
telescopes \citep[see ][]{2003whdw.conf..105H}. 
The MSST spectroscopic campaign was divided into two parts, the first of which 
(19--27/05/2002) was carried out at the Steward 
Observatory 2.3\,m on Kitt Peak, the Danish 1.54\,m at La Silla, and the 2.3\,m 
Advanced Technology Telescope at Siding Spring
Observatory. The second part (13--25/06/2002) took place about three 
weeks later, again at the Danish 1.54m at La Silla and at 
the 2.56\,m Nordic Optical Telescope at La Palma. Details are given 
in Table~1 of Paper~I. Spectra taken at ESO between June 13, 2002 and 
June 18, 2002 suffer from poor weather conditions resulting in so low S/N 
that they cannot be used for the spectral analysis. 

Due to the different instrumentation of the 
observatories, the spectral resolution and coverage are quite dissimilar (see
Table~\ref{tab_instruments}). Therefore the data sets of 
each telescope were analysed separately. During
the first part of the campaign only three nights of observation 
were secured at Siding Spring Observatory and only two nights at ESO.
The spectra obtained at Steward Observatory outnumber that of the former and are
of much better S/N. Hence the analysis of the first part of the campaign is
mainly based on the Steward data. The second part is dominated by observations
from the NOT as they outnumber the ESO spectra and are of better S/N than the
latter. Therefore the SSO and ESO spectra serve to check for 
consistency only (see Fig.~\ref{fig_dan_sso}). 
\begin{table}
\begin{normalsize}
\caption{Instrumentation of the used observatories}\label{tab_instruments}
\begin{center}
\begin{tabular}{lccr}
\hline\hline
Observatory & Resolution &$\lambda$
\hspace{.1cm}range & \# $\mathrm{spectra}$ \\
 & (\AA{}) & (\AA{}) & used\\
 \hline 
 \multicolumn{4}{l}{MSST: first half (19/05/2002--27/05/2002):}\\
Steward 2.3\,m & 1.8 & 3686-4534 & 2595 \\ 
Siding Spring 2.3\,m & 2.2 & 3647-5047 & 852 \\ 
Danish/ESO 1.54\,m &5.9 & 3648-5147 & 663\\
\hline 
\multicolumn{4}{l}{second half (18/06/2002--25/06/2002):}\\
Danish/ESO 1.54\,m &5.9 & 3648-5147 & 1562\\
NOT 2.56\,m & 9.3 & 3039-6669 & 3320 \\ \hline 
\end{tabular}
\end{center}
\end{normalsize}
\end{table}
The entire data set has already been analysed for radial velocity variations  
(Paper~I) and 20 modes have been detected with radial velocity
amplitudes between 0.8 and 15.4\,kms $^{-1}$.
In Paper I, we calculated the spectral windows for
the entire campaign as well as for the two halves.
These served as a guide to the aliasing that had to be
expected in the amplitude spectrum.
In order to get similar primers for the Steward and NOT
data sets exactly as they were used in the anaysis below,
this excersise has now been repeated for the two subsets separately.
Both window functions are very similar to each other, with a
resolution of about 2-3\,$\mu$Hz, and show the typical daily alias
patters (side lobes separated by 11\,$\mu$Hz) inherent to
single-site observations spread over several nights. In the next step 
the radial velocity data set was pre-whitened in order to get rid of the 
main peak and aliases induced by the dominant mode.

\begin{figure}
\begin{center}
\includegraphics [angle=270,scale=.3]{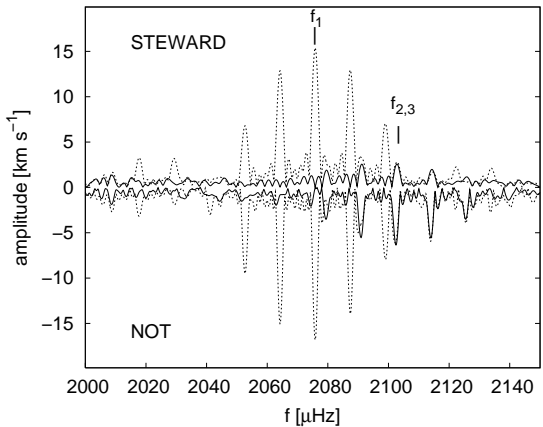}
\caption{\hspace*{0.5cm}Fourier Transform around the mode f1 (dashed line) of the radial velocities 
for the Steward data (upper part) and the NOT data (lower part, mirrored). The full drawn line showes
the results from pre-whitening for the strongest mode f1.}\label{fig_window_steward_not}
\end{center}
\end{figure}

The individual Discrete Fourier Transforms of the radial velocities 
with and without pre-whitening
for the Steward and the NOT campaigns in the vicinity of the peak of
the dominant mode are shown in Fig. \ref{fig_window_steward_not}. The amplitudes 
of the most important modes are given in Table 3.
\begin{figure*}[t]
\begin{center}
\includegraphics [angle=270,scale=.7] {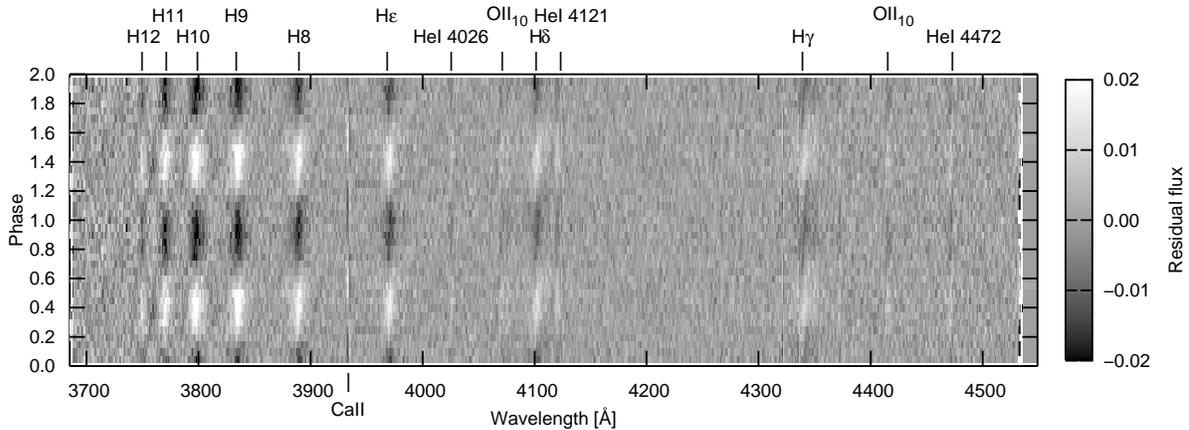}
\caption{\hspace*{0.5cm}Line profile variations of the phase binned Steward 
data for the strongest mode f1 (period P=481.74\,s). The CaII line is of interstellar nature. Its variation is due to the stellar radial velocity correction. Besides the strong Balmer, \hei\ and \Oii\ lines are found to vary.}\label{fig_var_steward}
\end{center}
\end{figure*}
The one and two day alias
peaks on each side of the main mode are obvious. As can be seen from Fig. 1 the amplitude 
of the dominant mode f1 is almost the same for both halves of the campaign. 

Pre-whitening the radial velocities for the strongest mode f1 (see Fig. \ref{fig_window_steward_not}) 
demonstrates that the unresolved modes f2 and f3 (with a separation of 0.66\,$\mu$Hz, at 2102\,$\mu$Hz) 
are unaffected by aliases of f1. But as we are using a phase binning technique it is not obvious if this 
statement is also valid for the changes in the line profile. 

It is evident from Fig. \ref{fig_window_steward_not} that the radial velocity amplitude of f2,3 is much
stronger in the NOT data than in the Steward data, an observation that
we will come back to when discussing the results of the quantitative spectral analysis
presented in the following. The same is true for the mode f4, which is situated in an area of the frequency 
spectrum where no aliasing effects of the dominant mode are expected (see Table 3). 

\begin{figure}
\begin{center}
\includegraphics [angle=270,scale=.6]
{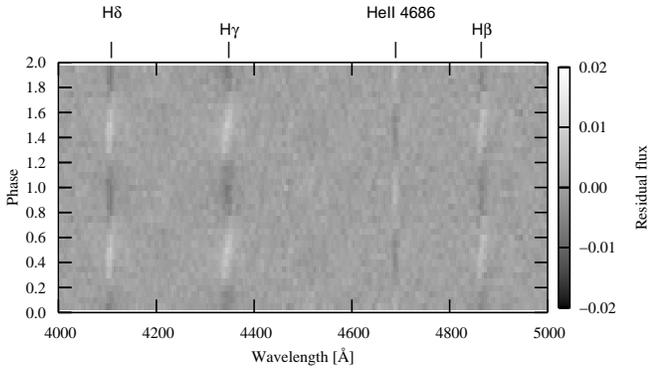}
\caption{\hspace*{0.5cm}Line profile variations of the phase binned NOT spectra 
for the strongest mode f1 (period P=481.74\,s). Note the presence of \heii\ 4686\,\AA varying in antiphase with the Balmer lines.}\label{fig_var_not}
\end{center}
\end{figure}

As individual spectra are too noisy for a quantitative spectral analysis in terms of detecting 
very small variations, they were combined in an appropriate way. 
For each spectrum the Doppler shift produced by the star itself and the heliocentric
correction was removed. The required values for each single spectrum have already been determined and used 
in Paper I. After that a continuum was fitted to the spectrum and a dispersion correction was applied.
To be able to detect tiny variations for any pre-chosen pulsation mode, 
we determined the phase of each individual spectrum (with respect to 
the same ephemeris zero point $t_0(HJD)=2452413.27166681$)
according to the selected pulsation period and co-added them accordingly. 
To this end a complete 
pulsation cycle was divided into twenty phase bins. 
In order to achieve the best possible 
result, we include a weighting procedure, in which every single spectrum is 
co-added according to its signal-to-noise-ratio.   
The quality of the co-added spectra improved significantly.
In Fig.~\ref{fig_var_steward} the phase dependent changes in the line profiles 
of the Steward 
data with respect to the mean spectrum are shown.
It is evident that all the observed H, \hei\ and \Oii\ lines vary in the same way 
with phase. Unfortunately there are no 
\heii\ lines present in the spectral range of the data from the 
Steward observatory. The Nordic Optical Telescope does give a broader wavelength
coverage at much lower spectral resolution. Fig.~\ref{fig_var_not} displays a
section of the NOT spectra covering \heii\ 4686\,\AA\ along with \hei\ 4471\,\AA\ 
and three Balmer lines. Unfortunately the spectral resolution is not high enough to
clearly detect the \hei\ 4471\,\AA\ , but the Balmer lines show the same behaviour
(see Fig.~\ref{fig_var_steward}). 
As expected the variations of the \heii\ line are
anti-correlated with those for the Balmer lines.




\section{Quantitative spectral analysis and atmospheric 
parameter variations}\label{sec:spec_analysis}

Effective temperatures ($T_{\rm eff}$), surface gravities ($\log{g}$), and helium
abundances ($y = N_\mathrm{He}/N_\mathrm{H}$) were determined by fitting
synthetic model spectra to all hydrogen and helium lines simultaneously, 
using a procedure developed by \citet{1999ApJ...517..399N}.
A grid of metal line-blanketed LTE model atmospheres with solar metal abundance
\citep{heber_2000_nlte_discu} as well as a grid of partially line-blanketed NLTE 
model atmospheres \citep{1997A&A...322..256N} was used.
LTE and NLTE atmospheric parameters differ slightly due to
systematic differences between these model grids.

\begin{figure}
\begin{center}
\includegraphics [scale=.45] {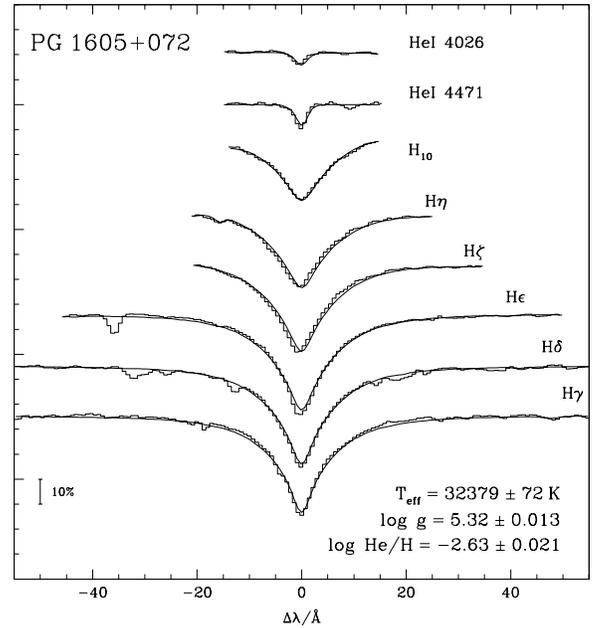}
\caption{\hspace*{0.5cm}Fit of the observed H/He lines from a Steward spectrum. The $\mathrm{H}_\epsilon$ core is affected by the interstellar Ca {\sc ii}, H line.
}\label{fig_fit_model}
\end{center}
\end{figure}

We derived the mean effective temperatures, gravities and helium abundances 
for each of the four data set (see Table \ref{Tab_f1_all}) as will be 
described in section 3.1 . When 
comparing the results one must
take into account that not only the spectra are of different resolution but that
also different sets of spectral lines had to be used due to limitations by
different wavelengths coverages. 
As can seen from Table \ref{Tab_f1_all} the formal  
statistical fitting errors are less than 70\,K, 0.015\,dex and 0.020\,dex,
respectively. As the differences between the results from different 
data sets and different model grids are much larger, we conclude that 
the formal errors are unrealistically small and 
the error budget is dominated by systematic errors both from observations as
well as
from model atmospheres. More realistic error estimates are 500\,{\rm K} for 
$T_{\rm eff}$, 0.05\,dex for $log\,g$, and 0.1\,dex for 
$log\,N(He)/N(H)$ \citep[for a more detailed discussion see][]
{2000A&A...363..198H,2006A&A...452..579O,2007A&A...464..299G}. Within these limits the results from the Steward, the NOT
and the SSO data sets agree very well with each other as well as with the
result of \citet{1999A&A...348L..25H}. The ESO spectra, however, give
significantly lower temperatures. 
However, the variation of the atmospheric parameters can be determined 
with much higher precision than their absolute values as the systematic errors 
cancel to a very large extend in a strictly differential analysis.\newline
In Fig.~\ref{fig_fit_model} a fit of the Steward spectrum for one phase bin
(weighted sum of more than 100 individual spectra) 
is displayed.   

We are now prepared to search for and analyse variations of the atmospheric
parameters, effective temperatures, surface gravities, and helium
abundances ($y = N_\mathrm{He}/N_\mathrm{H}$). As the dominant mode in radial
velocity is expected to also show the largest variation in temperature and
gravity, we start with this mode and
repeat the procedure for weaker modes. 

Note that the apparent variations of the atmospheric parameters are disk intergrated. 
The actual amplitudes at different positions on the stellar surface can be 
considerably larger depending on the type of mode. For a radial one disk-intergrated 
values should differ only slightly from the actual values on the surface. But for non-radial 
modes the importance of cancellation effects grows with rising degree l.


\subsection{The strongest mode}\label{sec:strong_mode}

\begin{figure}
\begin{center}
\includegraphics [scale=.5]{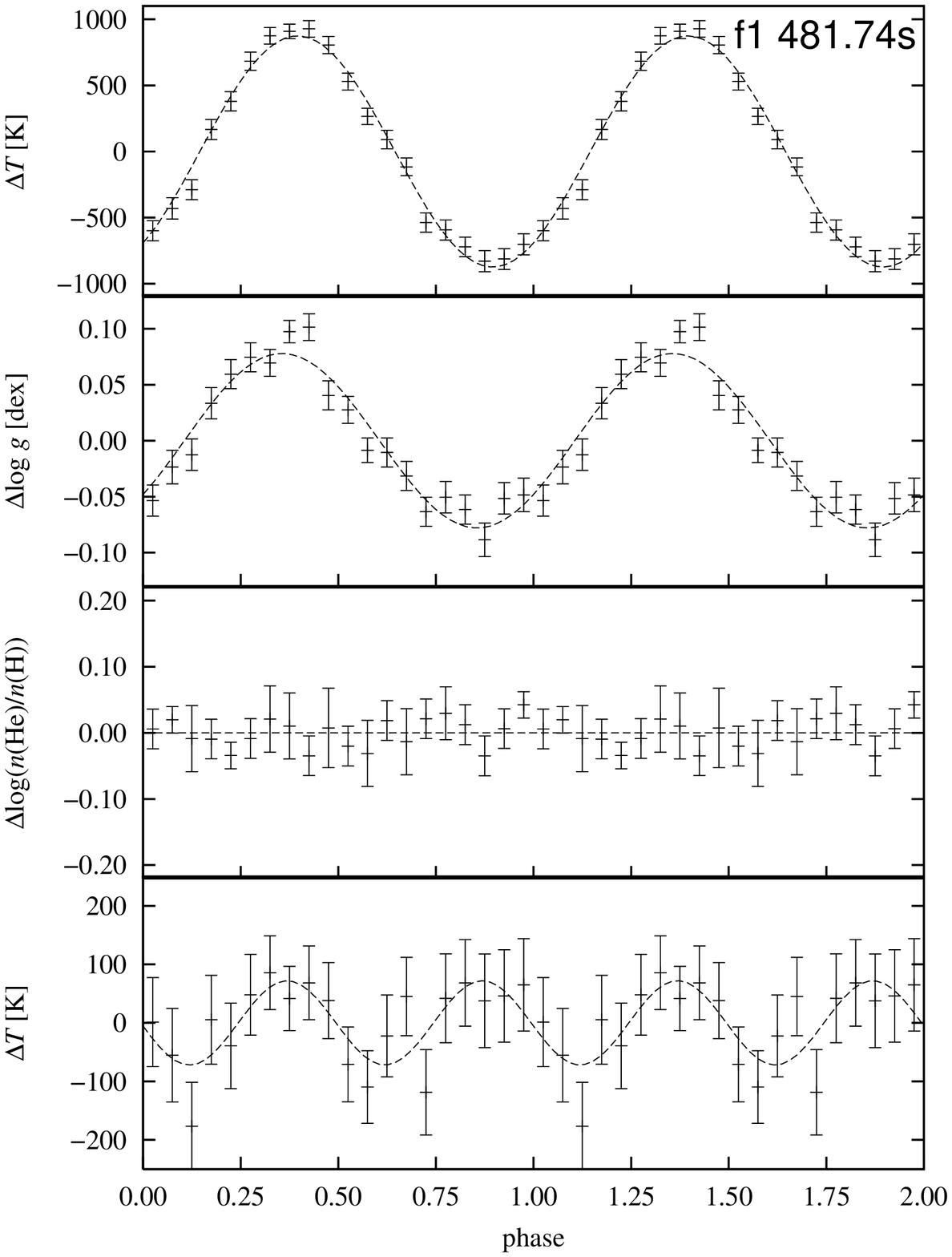}
\caption{\hspace*{0.5cm}The variations of the atmospheric parameters determined 
from the Steward data with statistical error bars and sine fit are shown. \newline
Upper panels: temperature and surface gravity with sine fits are shown.\newline 
Lower panels: He/H abundance and the temperature residuals (with a sine fit for the first harmonic) are shown.}\label{fig_atm_steward}
\end{center}
\end{figure}

In the first step we considered the mode with the largest radial velocity
amplitude and stacked the individual spectra accordingly into 20 phase bins.
We determined the variations of the three atmospheric parameters ($T_{\rm{eff}}$, 
$\log{g}$ and $\log{He/H}$-ratio) for every bin. 
Fig.~\ref{fig_atm_steward} shows the variations of the atmospheric parameters
for the Steward Observatory data. 
As can be seen the variations of $T_{\rm{eff}}$ and  
$\log{g}$ are sinusoidal. Therefore the pulsation semi-amplitude is determined 
by using a $\chi^2$ sine fitting procedure.

The temperature semi-amplitude is $874\,\mathrm{K}$, while the surface gravity 
variation is $\pm0.078\,\mathrm{dex}$. 
We also plotted the He/H abundance 
to make sure that it does not show periodical variations over a cycle. The remaining 
temperature residuals still show a periodicity (of $\pm72\,\mathrm{K}$) with frequency f10 
which was also already identified as the first harmonic in Paper I. As expected the amplitude 
ratio between the RVs (15.43\,\kms : 1.36\,\kms = 11.3) is similar to the ratio of the temperature 
amplitudes (873.7\,K : 72.0\,K = 12.1). Analogue we succeded to detect the first harmonic also in
the $\log{g}$ residuals with an amplitude of $\pm0.013\,\mathrm{dex}$, whereas the amplitude ratio 
is significant lower (0.078\,dex : 0.013\,dex = 6.0).

\begin{figure*}
\begin{center}
\includegraphics[angle=270,scale=.45]{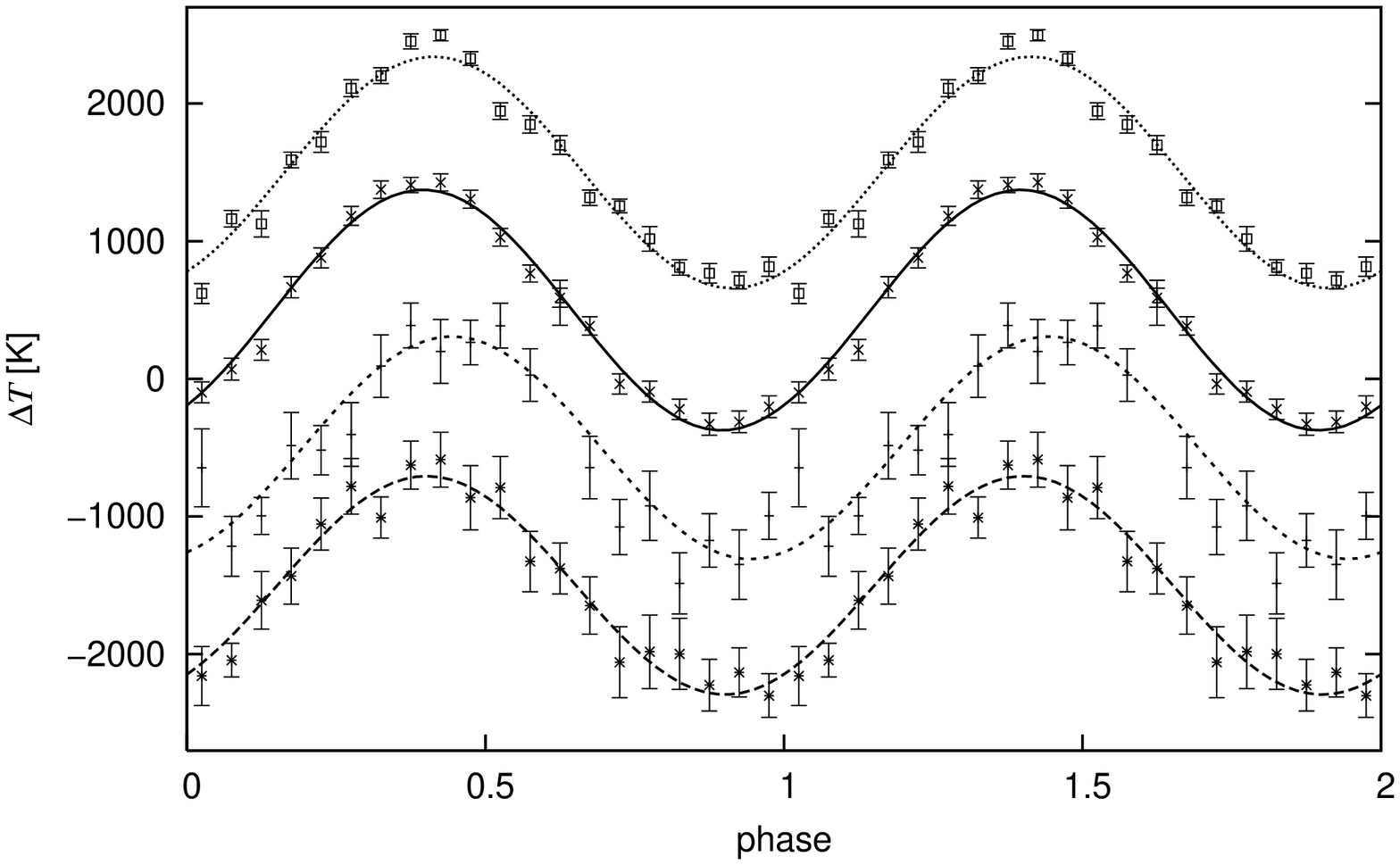}
\includegraphics[angle=270,scale=.45]{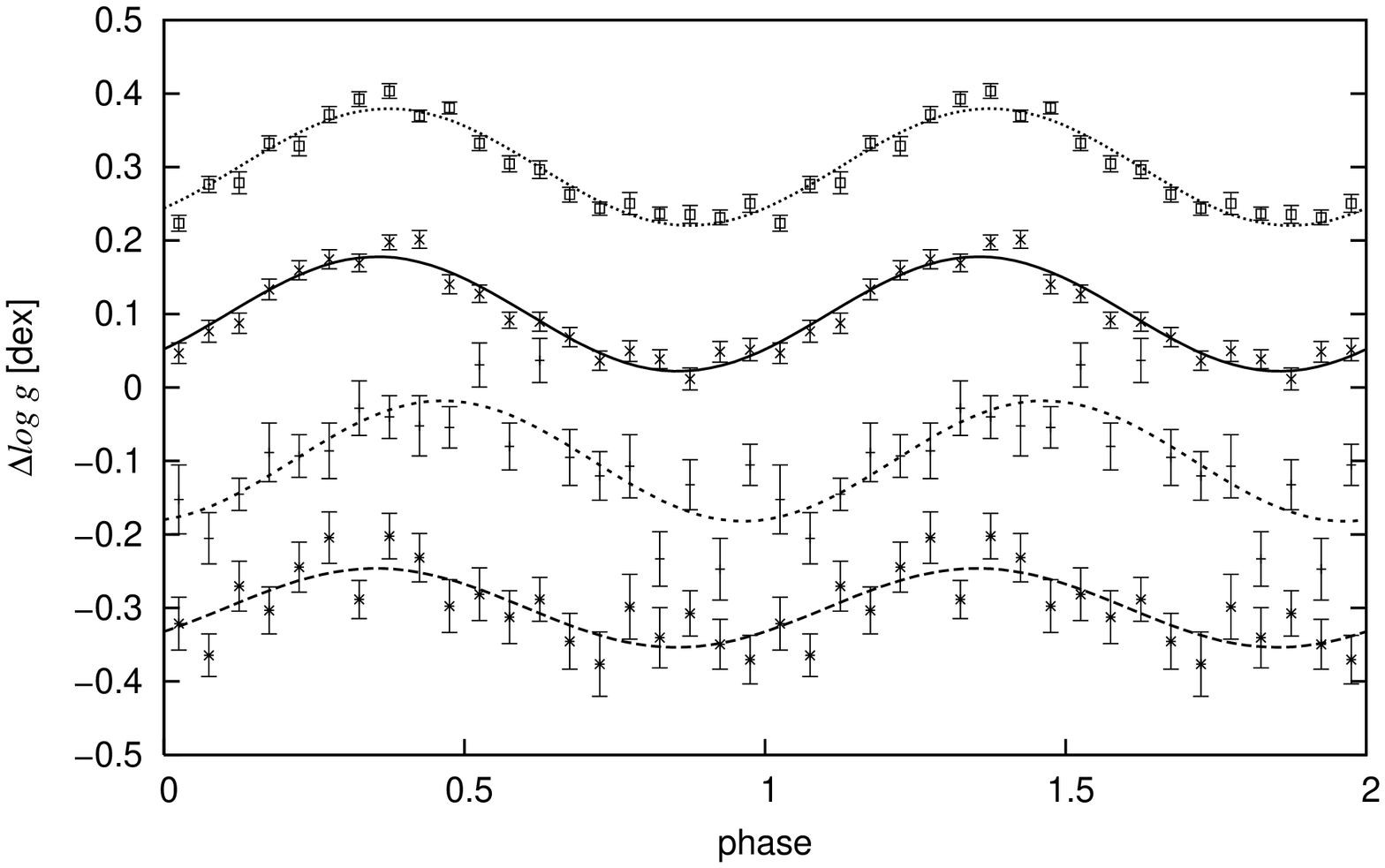}
\caption{\hspace*{0.5cm}Temperature and log g variations of the dominant mode f1 from all available data sets; from top to bottom:
NOT($\square$), Steward(x), SSO(+) and the ESO($\divideontimes)$. For clarity, the curves have been offset by 750\,K in $T_{\rm eff}$ and 0.1\,dex in gravity. 
The best fit sine curves are the dotted (NOT), full drawn (Steward), short dashed (SSO) and dashed (ESO) lines.}\label{fig_dan_sso}
\end{center}
\end{figure*}

Fig.~\ref{fig_dan_sso} compares the temperature and gravity variations of the strongest mode
f1 derived from all four data sets. The semi-amplitudes, derived from a sine fit as well, 
are consistent within error limits (see Table~\ref{Tab_f1_all}), except for the gravity amplitudes 
in the ESO spectra. This is hard to explain as the ESO observations overlap in time with 
those at the NOT. We conjecture that this is due to the limited number and lower quality of the 
ESO spectra.

\begin{table*}
\begin{center}
\begin{tabular}{lccccc}
\hline\hline
Observatory & $\overline{T_{\rm eff}}$ [K] & $\overline{\log{g}}$  [dex] & $\overline{\log{n_{\rm{He}}/n_{\rm{H}}}}$  [dex] & $\Delta T_{\rm eff}$ [K] & $\Delta\log{g}$ [dex]
\\ 
\hline
Steward & 31999.9$\pm$17.3 & 5.262$\pm$0.004 & -2.59$\pm$0.005 & 873.7$\pm$16.5 & 0.078$\pm$0.003\\ 
NOT     & 31927.9$\pm$27.1 & 5.299$\pm$0.004 & -2.58$\pm$0.007 & 840.9$\pm$26.0 & 0.079$\pm$0.004\\ 
SSO     & 32426.8$\pm$67.7 & 5.260$\pm$0.011 & -2.46$\pm$0.015 & 808.9$\pm$64.9 & 0.082$\pm$0.010\\ 
ESO     & 31437.7$\pm$31.5 & 5.224$\pm$0.008 & -2.58$\pm$0.011 & 793.7$\pm$30.2 & 0.054$\pm$0.008\\ 
\hline
\end{tabular}
\caption{The dominant mode f1: comparison of temperature and surface gravity 
mean values and semi-amplitudes from different data sets.}
\label{Tab_f1_all}
\end{center}
\end{table*} 


As the variations with phase of the atmospheric parameters for the dominant mode 
f1 have been 
measured in all data sets beyond doubt, we proceeded to search for and analyse 
weaker modes. We re-binned all spectra according to their phases with respect to
the frequency of a weaker mode and proceeded with all modes with decreasing
radial velocity amplitude (f2, f3 and so on). For these three modes with lower amplitudes 
(f2-f4) periodic variations of the atmospheric parameters were found in both
the Steward and the NOT data set.
As a test we applied the procedure to a random frequency
not detected in velocity or light. Periodic variation of the atmospheric 
parameters were found neither in the Steward nor in the NOT data 
indicating that the dominant mode produces sufficient self-cancellation for its 
influence to be annihilated. 

\subsection{Cleaning the spectra}\label{sec:prewhitening}

As is clearly seen from Fig. 1, the main mode produces alias peaks close 
to the location of f2 and f3. For f4 onwards the influence is much 
smaller, but due to the high amplitude of the main mode the influence 
can still be significant. Unlike the case for the photometry and radial 
velocity data, which can be cleaned using a standard pre-whitening 
procedure, a more sophisticated approach is required to remove the 
influence of the main mode from the spectroscopic data. Hence we have 
implemented a cleaning procedure based on our synthetic model spectra. 

\begin{figure*}
\begin{center}
\includegraphics [scale=.5]{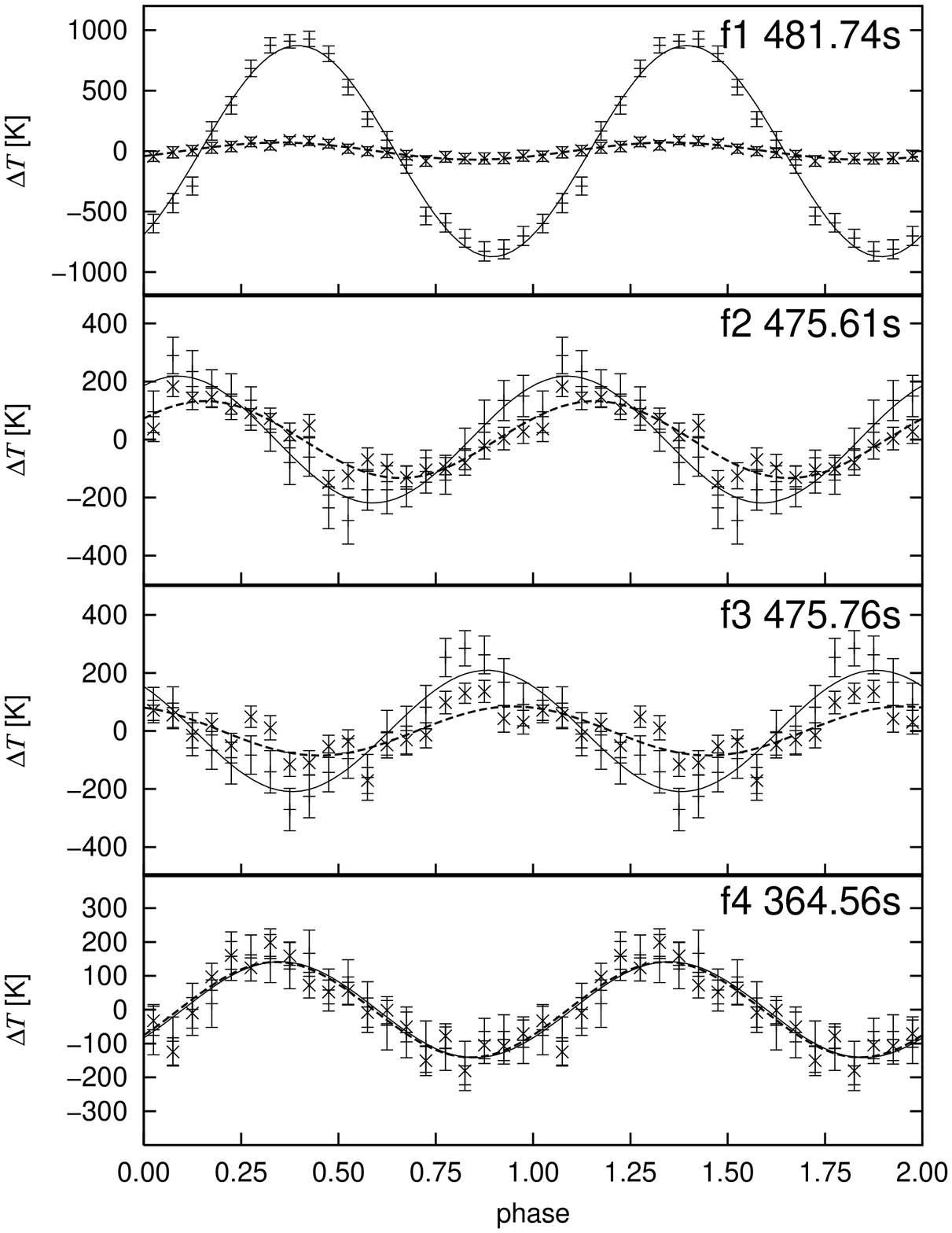}
\includegraphics [scale=.5]{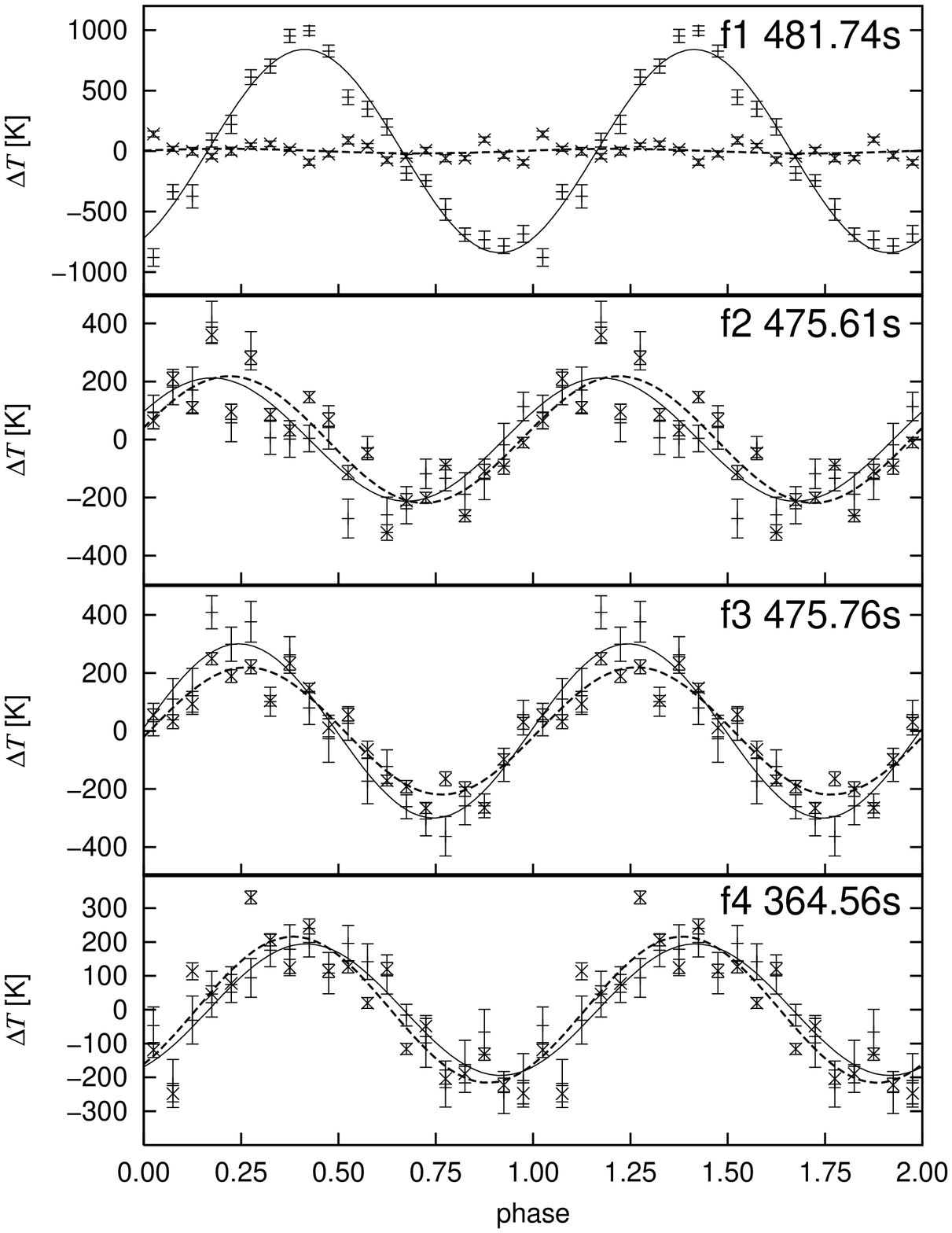}
\includegraphics [scale=.5]{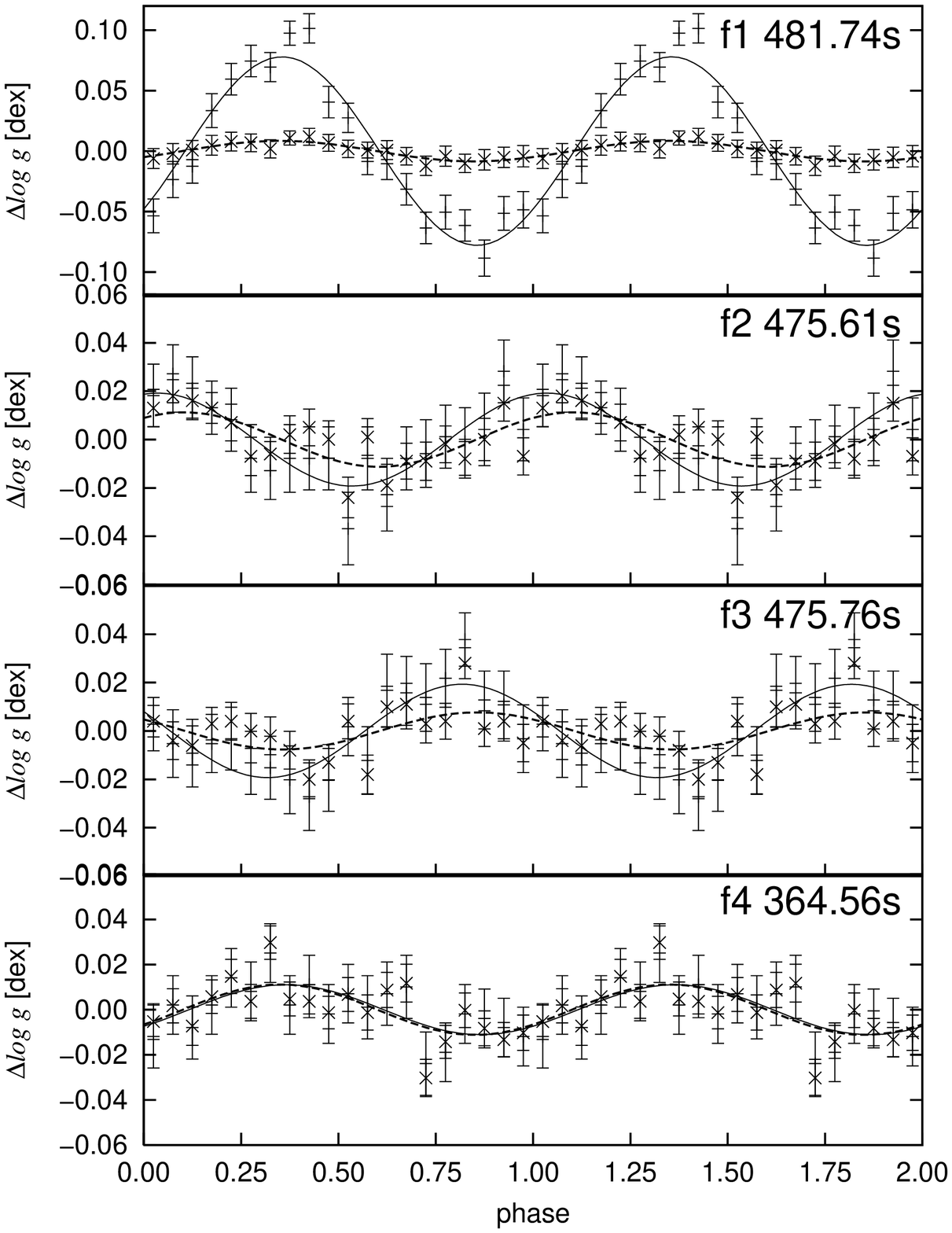}
\includegraphics [scale=.5]{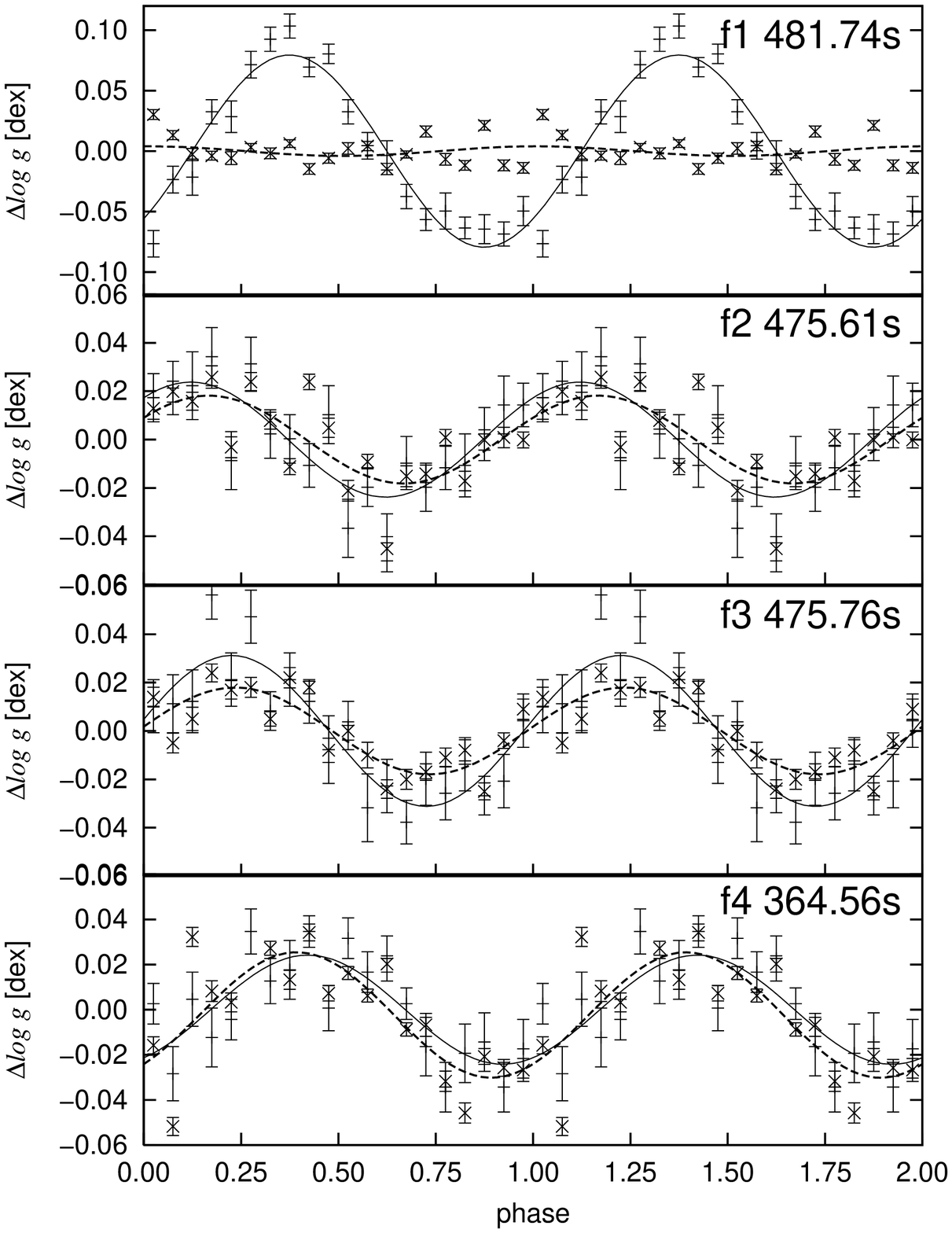}
\caption{\hspace*{0.5cm}\textit{left hand side}: Steward spectra: variations of the temperature (top) and the log g (bottom) for the four strongest 
modes before (+) and after (x) the cleaning procedure with the statistical error bars. The best fit sine curves are the full drawn (without the cleaning) and dashed (after the cleaning) lines.\newline
\hspace*{0.5cm}\textit{right hand side}: same for the NOT spectra.}\label{fig_f124}
\end{center}
\end{figure*}

First we calculated synthetic spectra for every phase bin of the dominant pulsation
mode f1. Then they were summed up to create the mean synthetic spectrum for f1, which was 
subtracted from each phase bin spectrum to form the cleaning function. 
This correction was applied to all of the $\sim 9000$ individual spectra by
subtracting the cleaning function for the corresponding phase of
the dominant mode. The cleaned spectra were then summed into 20 phase
bins for the corresponding frequencies of lower amplitude modes (from f2 onwards)
and were analysed for the atmospheric parameter variations as described
above.
As a test we applied the procedure to the dominant mode f1 as well. 
The residual amplitudes are listed in Table 3 and shown in
Fig.~\ref{fig_f124}. As can be seen, no significant residuals remain for the NOT data, while 
small residuals remain in particular in the
Steward spectra, which may be basically due to the 1st harmonic mode f10 (see section \ref{sec:strong_mode}) and the limitations of the model spectra as we assume e.g. a homogenous temperature distribution on the surface.

\begin{table*}
\begin{center}
\begin{tabular}{cccccccc}
\hline\hline
Name & Period & f & v &$\Delta T_{\rm eff}$ &  $\Delta T_{\rm eff}$ & 
$\Delta\log{g}$ & $\Delta\log{g}$ \\
& (s) & ($\mu$Hz) & $kms^{-1}$ & (K) & (K) & (dex) & (dex) \\
&     &           &            & no cleaning & cleaning & no cleaning & cleaning \\ 
\hline
\multicolumn{8}{l}{Steward Observatory}\\
f1 & 481.74 & 2075.80 & 15.429 & 873.7$\pm$16.5 &  70.5$\pm$4.7   & 0.078$\pm$0.003 & 0.008$\pm$0.001\\ 
f2 & 475.61 & 2102.55 & 5.372  & 218.5$\pm$15.3 &  132.6$\pm$12.6 & 0.019$\pm$0.002 & 0.011$\pm$0.002\\ 
f3 & 475.76 & 2101.91 & 2.971  & 209.1$\pm$18.3 &  84.5$\pm$17.5  & 0.019$\pm$0.002 & 0.008$\pm$0.002\\ 
f4 & 364.56 & 2743.01 & 2.497  & 141.8$\pm$11.2 &  140.9$\pm$14.6 & 0.011$\pm$0.002 & 0.011$\pm$0.002\\ 
f5 & 503.55 & 1985.89 & 2.474  &        -       &  117.9$\pm$10.3 &         -       & 0.014$\pm$0.002 \\ 
f6 & 528.71 & 1891.41 & 2.322  &        -       &  87.7$\pm$15.0  &         -       & 0.009$\pm$0.002\\ 
f7 & 361.86 & 2763.50 & 2.121  &        -       &  136.8$\pm$10.5 &         -       & 0.013$\pm$0.003\\ 
f8 & 246.20 & 4061.70 & 1.777  &        -       &  88.1$\pm$18.3  &         -       & 0.021$\pm$0.002\\ 
\hline
\multicolumn{8}{l}{NOT Observatory}\\
f1 & 481.74 & 2075.80 & 15.429 &  840.9$\pm$26.0 &  18.9$\pm$21.3  & 0.079$\pm$0.004 & 0.000$\pm$0.003\\ 
f2 & 475.61 & 2102.55 & 5.372  &  212.6$\pm$31.9 &  218.4$\pm$28.3 & 0.024$\pm$0.003 & 0.018$\pm$0.003\\ 
f3 & 475.76 & 2101.91 & 2.971  &  300.3$\pm$24.4 &  219.5$\pm$18.3 & 0.031$\pm$0.003 & 0.018$\pm$0.002\\ 
f4 & 364.56 & 2743.01 & 2.497  &  194.5$\pm$19.8 &  215.8$\pm$27.1 & 0.024$\pm$0.003 & 0.031$\pm$0.004\\ 
\hline
\end{tabular}
\caption{Semi-amplitudes of the temperature and gravity variations before and after 
cleaning for the four strongest modes derived from Steward and NOT data. Also the statistical error from the $\chi^2$ sine fitting is listed. Periods, frequencies and radial velocity amplitudes are taken from Paper I and therefore calculated for the whole MSST data run.}
\end{center}\label{tab_pre}
\end{table*}

 
An important improvement achieved by the cleaning is that the statistical errors 
of the individual atmospheric parameters derived for each 
phase bin decrease by a factor of 2, allowing the detection of even 
weaker variations in temperature ($\sim80\rm{K}$) and gravity ($\sim0.01\rm{dex}$), 
as demonstrated below. 

The amplitudes of the atmospheric parameter variation for f4 are 
unaffected by the cleaning procedure. However the amplitudes derived from the
Steward data are significantly lower than those from the NOT data. This is consistent 
with the lower RV amplitude in the Steward data (see Sect. \ref{sec:data}). 

Unlike f4, the amplitudes for f2 and f3 of the Steward data decreased after cleaning. 
The same holds for f3 in the NOT data, while f2 is almost unaffected by
cleaning. Note that f2 and f3 are so close in frequency that they are not resolved in either data set. 

At this point we would like to refer again to Fig.\ref{fig_window_steward_not}; the amplitude in radial velocity
for f2,3 is much higher in the NOT data than in the Steward data set and the same holds for f4. The differences 
in the detected temperature and gravity variations between the Steward and the NOT data sets are consistent 
with the radial velocity amplitudes and therefore plausible.


\subsection{The weak pulsations f5--f11}\label{sec:subordinate}

\begin{figure*}
\begin{center}
\includegraphics [scale=.5]{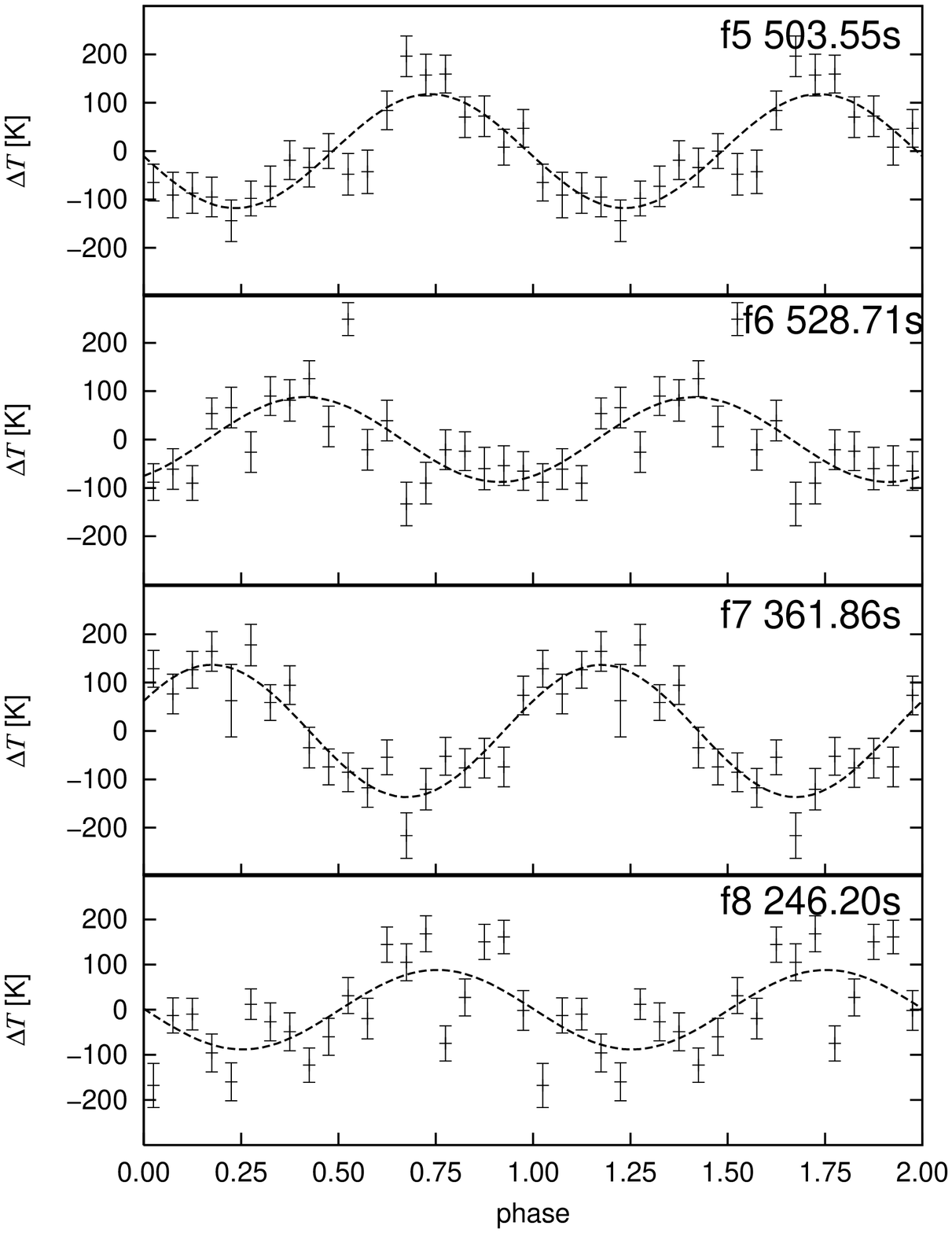}
\includegraphics [scale=.5]{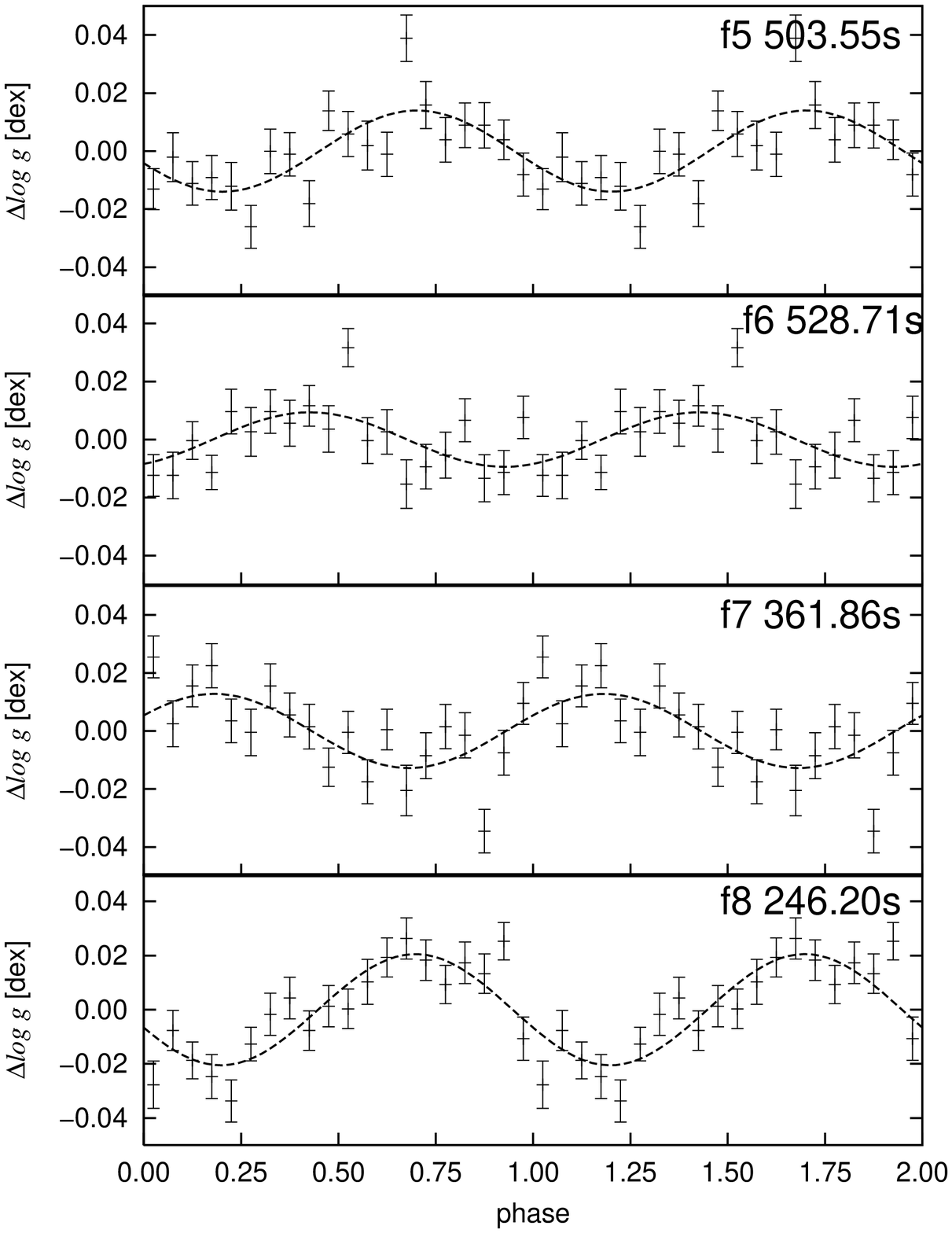}
\caption{\hspace*{0.5cm}Variations of the temperature(left) and the log g(right) for the modes 
f5, f6, f7 and f8 after the cleaning procedure with sine fits and 
statistical error bars derived for the Steward data.}\label{fig_weakmodes}
\end{center}
\end{figure*}

Without the cleaning we were able to detect temperature and gravity variations only for those four frequencies with the  
largest radial velocity amplitudes. Since the statistical errors of atmospheric 
parameters are significantly reduced by the cleaning procedure, it is
worthwhile to search for additional frequencies. 
The Steward spectra are suited best because they have the highest spectral 
resolution, S/N, and the second largest number of spectra. Indeed, we detected 
weaker modes (see Fig.\ref{fig_weakmodes}). The semi-amplitudes determined 
again by $\chi^2$ fitting of a sine function are listed in Table 3. 
For the mode f9 the errors of our data points ($\delta T_{\rm eff}=40\,\rm{K},\delta\log{g}=0.008\,\rm{dex}$) 
are of the same order of magnitude as the amplitude ($\Delta T_{\rm eff}=35\,\rm{K},\Delta\log{g}=0.003\,\rm{dex}$) 
and the detection of variations must be regarded as marginal.\\
Nevertheless, this demonstrates that it is possible to 
reveal variations of atmospheric parameters of modes with radial velocity
variations down to 2 \kms. 

Frequencies f5 and f6 are isolated and therefore well resolved.
Temperature variations are found at amplitudes of 110\,K and 90\,K, respectively 
while gravity variations of 0.014\,dex and 0.008\,dex are detected. 

Frequencies f7, f9, and f11 are close together and therefore unresolved. While 
amplitude variations of atmospheric parameters are found for f7, the variations
for f9 and f11 are below the detection limit. 

f8 and f10 are combination frequencies involving f1 (f1+f5 and 2$\times$f1,
respectively). For f8 the variation of $T_{\mathrm{eff}}$ is detected 
marginally only, whereas the gravity variation is pronounced (see Fig.\ref{fig_weakmodes} and Table 3). 
f10 was already detected in the residuals of the sine fit to f1 as discussed in Section 3.1 
(see also Fig. \ref{fig_atm_steward}).

\section{Phase lags}\label{sec:phase_lags}

In order to characterise pulsation modes it is important to investigate phase
lags between temperature, gravity and radial velocity variations as well. 
These phase lags can be used to determine the deviations from a linear adiabatic system. 

\subsection{Temperature versus radial velocity}
\begin{figure}
\begin{center}
\includegraphics [scale=0.5]{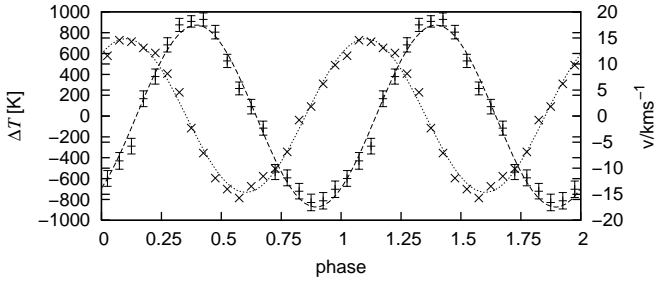}
\caption{\hspace*{0.5cm}Radial velocity (x) and 
the temperature variation (+) of the dominant mode f1 derived from the Steward 
data. The dashed line is the best fit sine curve for the $T_{\rm eff}$ variations; 
the dotted line is the best sine curve for the radial velocity variation.}\label{fig_rad_teff}
\end{center}
\end{figure}

The radial velocities of individual spectra were taken from Paper I. The amplitudes 
of their variations were determined in the same
way as the temperature and gravity variations using the phase binning technique and cleaning 
described in Sect 2. Phase lags between the variation of radial velocity and temperature can be derived
directly by comparing the phases of the maxima of the $T_{\rm eff}$ and RV curves.
\begin{table*}
\begin{center}
\begin{tabular}{lcccccccc}
\hline\hline
& \multicolumn{4}{c}{Steward} & \multicolumn{4}{c}{NOT} \\
& $\Delta\varphi (T_{\rm{eff}}-RV)$ & $\Delta\varphi(T_{\rm{eff}}-\log{g})$ && $\Delta\varphi (\log{g}-RV)$ & $\Delta\varphi (T_{\rm{eff}}-RV)$ & $\Delta\varphi(T_{\rm{eff}}-\log{g})$ && $\Delta\varphi (\log{g}-RV)$ \\
Name & cleaning & no cleaning & cleaning & cleaning & cleaning & no cleaning & cleaning & cleaning\\
\hline
f1 & +0.293 & +0.040 & -      & +0.253 & +0.311 & +0.039 & -      & +0.272 \\
f2 & +0.289 & +0.054 & +0.054 & +0.235 & +0.351 & +0.054 & +0.053 & +0.300 \\
f3 & -0.131 & +0.063 & +0.193 & -0.275 & +0.184 & +0.019 & +0.031 & +0.153 \\
f4 & +0.260 & -0.016 & -0.011 & +0.271 & +0.280 & -0.007 & +0.007 & +0.298 \\
f5 & +0.268 & -      & +0.034 & +0.234 & -      & -      & -      & -      \\
f6 & +0.219 & -      & -0.013 & +0.231 & -      & -      & -      & -      \\
f7 & +0.364 & -      & -0.006 & +0.370 & -      & -      & -      & -      \\
f8 & +0.238 & -      & +0.048 & +0.182 & -      & -      & -      & -      \\
\hline
\end{tabular}
\caption{Phase lags $\Delta\varphi$ (in units of $2\pi$) between $T_{\rm eff}$, radial velocity and $\log{g}$ derived from the Steward and NOT data before and after the cleaning. Positive values mean that the maximum of the radial velocity or gravity curve occurs before that of $T_{\rm eff}$ or $\log{g}$. The values for the radial velocities are taken from Paper I.}
\end{center}\label{tab_phaselag_TRV}
\end{table*}
The phase lag between the temperature and the radial velocity variation for the dominant mode is displayed in 
Fig.~\ref{fig_rad_teff}.
In order to determine the phase lags of the weaker modes we used the temperature curves after cleaning. 
We just measure the phase lags for two sine functions. 
The formal fitting errors are unrealistically small. Systematic errors are probably more important 
but hard to quantify. As we have two independant measurements at hand (Steward, NOT) we can use them to 
estimate the error. As f2/f3 are not suitable due to blending, we obtain an error estimate of 
$\pm 0.025$ from f1 and f4.

The results for all the recovered modes are shown in Table 4. 

It is worth noting that for all recovered modes exept f3 the radial velocity variation reaches its maximum before the temperature variation which means throughout positive values for the phase lags. 
Furthermore, all the values of the phase lags lie around $0.25$($\triangleq\frac{\pi}{2}$) with small but significant deviations. This is the value we would expect for a completely adiabatic p-mode pulsation. But as a real star is a non-adiabatic system due to its radiation of light, such deviations are to be expected (see Table 4). 

For the dominant mode f1 the phase lag between $T_{\rm{eff}}$ and $log{g}$ is 0.3 (see Table 4) indicating 
that the temperature is highest shortly after the radius is smallest. 


At this point we would like to emphasise our results for f3. Comparing the temperature variation plots for f3 in Figure \ref{fig_f124} it is obvious that the phase of the two sine functions without cleaning differs by a value of $0.5$ ($\triangleq\pi$) in phase between the Steward and the NOT results. At the same time, the phases of the radial velocity curves 
between the two main data sets differ by about $0.25$ ($\triangleq\frac{\pi}{2}$). After the cleaning the situation 
does not improve much, as we still meausure a phase lag of $+0.315$ for temperature and $+0.428$ for gravity between 
the Steward and the NOT data.

If real, this would indicate that a phase jump occurred between the two observing runs. 
However since f3 is unresolved from f2 (see Fig. \ref{fig_window_steward_not}) this phenomenon could well be 
an artifact. It is quite unfortunate that we could not combine Steward and NOT data in the quantitative analysis, because the 38\,days 
coverage should have given us the required time resolution.\newline 

\subsection{Temperature versus gravity}

As before the phase lags between the variation of temperature and gravity can be derived
directly by comparing the phases of the maxima of the $T_{\rm eff}$ and $\log{g}$ curves. 
In Table 4 we give the phase lags with and without cleaning.
The result is that the phase lags between $T_{\rm eff}$ and $\log{g}$ are rather small. 

The phase lags between $T_{\rm eff}$ and $\log{g}$ are insensitive to the cleaning procedure for f2. Within the maximum error limits of half the bin size (i.e. 0.025) the phase lags found in the Steward spectra are consistent with those from NOT spectra. 

For f3 the situation is very different as the phase lags show a rather large discrepancy between the two data sets. 
Referring again to Fig.\ref{fig_window_steward_not} and Fig.\ref{fig_f124} it is obvious that the mode f3 seems to change in both, phase and amplitude between the two parts of the campaign. Another point is that the peak of the unresolved modes f2/f3 is much higher in the NOT data than in the Steward data. If we take also into account the detection of a much higher amplitude in $T_{\rm eff}$ and $\log{g}$ for f3 after the cleaning (see Table 3 and Fig.\ref{fig_f124}), it is more likely that the phase lag for f3 from the NOT data constitute a more reliable measurement in this case.

\subsection{Radial velocity versus gravity}

Finally, we consider the phase lags between the radial velocity and the gravity. As they are both produced by the 
movement of the stellar surface, they are supposed to have a strict relationship. 
I.e. the g variation should be inline with the derivative of the radial velocity curve. Hence we expect a 
phase lag of $\frac{\pi}{2}$ between the RV curve and the $\log{g}$ curve.

We apply the same fitting procedure as before. The results are shown in Table 4.

Alternatively we can calculate $\Delta\varphi (\log{g}-RV)$ from 
$\Delta\varphi(T_{\rm{eff}}-RV)-\Delta\varphi(T_{\rm{eff}}-\log{g})$. The small difference between the two methods 
indicates the consistancy of our measurements.


For the Steward data most values seem to be in agreement with the theory within 0.02 consistent with 
our error estimate in Sect. 4.1, except for frequencies f3, f7 and f8. The difficulty with f3 have already been discussed (see above) and the same may hold for f7 as it also unresolved from f9, wereas frequency f8 is close to the detection limit (see Sect.\ref{sec:spec_analysis}). 
While the phase lags $\Delta\varphi(\log{g}-RV)$ are consistent with expectation (0.25) the phase lags from NOT data 
show a systematic offset of $\sim0.04$. 
Overall the phase lags from Steward data appear to be more reliable.

\section{Discussion}\label{sec:discussion}
The quantitative spectral analysis of about 9000 spectra of the pulsating 
subdwarf B star PG~1605+072 obtained in the context of the 
MSST campaign allowed us
to detect line profile variations of those four modes that show the
strongest radial velocity variations, as well as for the first harmonic of the 
dominant mode. The data set was obtained at four 
observatories in two observing periods in May 2002 and June 2002
separated by about three weeks. Spectra of the first half were predominantly 
taken at the Steward 2.3m telescope at KPNO, while those of the second half 
were mostly from the Nordic Optical Telescope. Spectra taken at the Siding
spring observatory and at ESO were used for a consistency check only.   

Using a new cleaning technique we
obtained more precise results, which allowed us to detect tiny line profile 
variations for four additional modes, including a combination frquency, 
with radial velocity variation as small as
1.8\,$\mathrm{kms^{-1}}$. Variations of the effective temperature and gravity are
sinusoidal and with semi-amplitudes ranging from 
90\,K to 870\,K in effective temperature and 
from 0.009\,dex to 0.078\,dex in gravity. We find evidence for amplitude
variations between the first and the second half of the MSST campaign
for f2/f3 and the f4 mode but not for f1. As f2/f3 are unresolved this could be
due to beating. The analysis of the contemporary MSST photometry indicates that 
f4 could actually be split in two modes (Schuh et al., 2007 in prep.) and the amplitude changes found 
could be as well caused by beating (Paper III).

We also measured the phase shift between the temperature, the gravity and the
radial velocity curves. Effective temperature and gravity show only a small phase lag. 
This is not the case for the phase shifts between radial velocity and
temperature. All phase shifts except for f3 in the Steward data lie around $\frac{\pi}{2}$ 
(between 0.20 and 0.30 in units of $2\pi$), which is expected for adiabatic pulsations, 
while f7 shows a somewhat larger phase of 0.37. Most notably, however, for the dominant mode 
the maximum of the star's temperature occurs shortly after the minimal radius.

\citet{2005A&A...442.1015K} measured radial velocity and flux variations
from far-UV time-resolved spectroscopy and determined a phase shift of
$\pi$ between the maxima of radius and flux of the strongest mode. 
As the flux variation is mainly
caused by temperature variations and maximum radius occurs at zero velocity,
this translates into a phase shift of $\pi/2$ between the temperature and radial
velocity variations, in agreement with our results.

The behaviour of the mode f3 poses problems. The amplitudes as well as the
phasing of the temperature, gravity and radial velocity variations are 
sensitive to the cleaning procedure. Moreover, there seems to be a phase 
jump between the two halves of the observing campaigns unlike for any other 
mode. Phase variations have been observed in the V361~Hya star HS~1824+5745 
\citep{2006MNRAS.369.1529R} but remained unexplained. However, 
as the frequency of f3 is very close to that of f2 they are unresolved at least in terms of
radial velocity. f3 may be strongly influenced by f2, which renders our results for f3 dubious.

Variations of the atmospheric parameters have already been derived by 
\cite{2003MNRAS.340..856O} from variations of Balmer line indices. The
measurements were based on time resolved spectroscopy carried out in 
1999 and 2000. The same grid of model atmospheres as used here was employed 
to derive effective temperatures and gravities from Balmer line indices.
Unlike in this paper, effective temperature and gravities were determined 
independantly. After measuring T$_{\rm eff}$ with a fixed gravity, the adopted 
gravities were determined keeping T$_{\rm eff}$ fixed.

The radial velocity amplitudes of individual modes change dramatically 
from year to year. The radial velocity variations of the dominant mode f1 
at 2075\,$\mue$Hz during the MSST campaign for instance increase 
from 2.4\,\kms\ in 1999 to 
4.27\,\kms\ in 2000 to 15.4\,\kms\ in 2002. Therefore it might be misleading to 
compare the amplitudes of temperature and gravity variations at different 
epochs. We would expect the temperature and gravity variations to scale 
with the amplitude of the radial velocity curve, which is indeed evident. 
The T$_{\rm eff}$/$\log{g}$ variations for 1999 and 2000 are much smaller 
than in 2002 as were the RV amplitudes \citep{2000ApJ...537L..53O}. 

\cite{2004A&A...419..685T} were the first to derive temperature and gravity
variation for a pulsating sdB star from fitting synthetic line profiles to
time-series spectroscopy. They detected line profile variations
in PG~1325$+$101 and derived temperature and gravity variation for the dominant
mode using the same grid of model atmospheres as used here. The amplitudes 
of the radial velocity, T$_{\rm eff}$ and $\log{g}$ for PG~1325$+$101 
(12.3\,\kms, 610\,K, 0.051\,dex, respectively) are somewhat smaller than for 
PG~1605$+$072 (15.4\,\kms, 857\,K, 0.079\,dex). Moreover, 
\cite{2004A&A...419..685T} measured 
phase lags between temperature, gravity and radial velocity for PG~1325$+$101
and found exactly the same values as we do for PG~1605$+$072 
($\Delta\phi$(T$_{\rm eff}-RV$)=+0.29 and 
$\Delta\phi$(T$_{\rm eff}-\log{g}$)=+0.04).

\cite{2004A&A...419..685T} assumed that the dominant mode in PG~1325$+$101
is a radial one and calculated the amplitude of the radius variation and the
amplitude of the pulsational acceleration. We apply the same procedure to 
PG~1605$+$072 and find that the variation of the radius is 0.9\%
and the pulsational acceleration is 0.292\,kms$^{-2}$ corresponding to variations
of surface gravity by $\Delta\log{g}$ = 0.008\,dex and 0.072\,dex, respectively.
As in the case of PG~1325$+$101 the pulsational acceleration is far more
important than the change in radius. The predicted change in gravity
is remarkably close to the measured value (0.078\,dex). However, 
we are reluctant to conclude that f1 is a radial mode, because the temperature 
variations have to be matched as well, which needs further detailed modelling. 

The MSST campaign has also obtained an unprecedented set of time resolved 
photometry, which will be presented and analysed in a forthcoming Paper~III. 
A small part of the MSST photometry has been used by \citet{2005ApJ...622.1068P}
to demonstrate that there is no stochastic mechanism exciting the 
oscillations in the subdwarf B star PG 1605+072. 
The full set of observations (radial velocity curve from Paper~I, temperature
and gravity variations from this paper and the light curve analyses) will then 
form a sound basis for astroseismology. 

For this purpose the modes have to be classified first, i.e. the 
pulsational ''quantum'' numbers have to be determined. 
This needs appropriate modelling. Sophisticated 
linear, adiabatic and non-adiabatic models for non-radial pulsations are available and 
have been successfully applied to match model prediction to the  
the observed periods from light curve analysis (see e.g. Charpinet et al. 2006).
 The case of 
PG~1605$+$072 is more complicated because rotation may play an important role.
Kawaler (1999) suggested a model that fitted the five strongest modes if the star
rotates rapidly at 130\,\kms. Heber et al. (1999) backed this up by 
measuring line broadening of 39\,\kms and interpreted this as 
the projected rotational velocity assuming that broadening due to pulsational
motions are negligible. Since it is now clear that pulsational motions add significantly 
to the line broadening, the projected rotational velocity is smaller than anticipated. 

Nevertheless, pulsational models have to account for rotation as well. A numerical 
code to model line profile variations due to adiabatic non-radial
pulsations for rapidly rotating early type stars has been developed by \citet{Townsend_thesis} 
and successfully applied to explain the pulsations of Be stars 
\citep{2003A&A...411..229R, 2003A&A...411..181M}. 
This code, named BRUCE, is well suited to model the pulsations of 
PG~1605$+$072 as well. It has already been used to calculate synthetic photometry 
for non-radial pulsations in subdwarf B stars by \cite{2004A&A...428..209R}. 
We shall apply this approach to model the light curve, the 
radial velocity curve as well as the temperature/gravity variations of PG1605+072 
in a forthcoming paper (Paper~IV).

\acknowledgements{Our thanks go to M. Billeres, E. M. Green, H. Kjeldsen, T. Mauch and V. Woolf for their observing efforts in the 
MSST campaign. S. J. O'Toole is supported by PPARC grant PPC/C000552/1.}


\bibliography{references}
\bibliographystyle{aa}

\end{document}